\pdfoutput=1
\documentclass{iau} 
\usepackage{graphicx}
\usepackage{natbib}
\usepackage{aas_macros}
\usepackage{hyperref}

\title[Life Beyond PTF] %% give here short title %%
{Life Beyond PTF}

\author[Eric C. Bellm]   %% give here short author list %%
{Eric C. Bellm$^1$}

\affiliation{$^1$Department of Astronomy, University of Washington, \\  Box 351580,  Seattle, WA 98195
\\ email: {\tt ecbellm@uw.edu}} 

\pubyear{2018}
\volume{339}  %% insert here IAU Symposium No.
\jname{Southern Horizons in Time Domain Astronomy}
\editors{E. Griffin, ed.}
\begin{document}

\maketitle

\begin{abstract}
In March 2017, the Intermediate Palomar Transient Factory (iPTF) ceased operations.  
I take this occasion to review the scientific returns from iPTF and its predecessor survey, the Palomar Transient Factory (PTF), and to summarize the lessons learned.  
Succeeding iPTF on the Palomar Observatory 48-inch Schmidt telescope is the Zwicky Transient Facility (ZTF), a new survey with an order of magnitude faster survey speed that is now being commissioned.  
I describe the design and scientific rationale for ZTF.  
ZTF is prototyping new alert stream technologies being explored by the Large Synoptic Survey Telescope (LSST) to  distribute millions of transient alerts per night to downstream science users.  
I describe the design of the alert system and discuss it in the context of the wider LSST and community broker ecosystem.  

\end{abstract}

\firstsection % if your document starts with a section,
\section{The PTF and iPTF Surveys}

The Palomar Transient Factory survey  \citep{Law:09:PTFOverview} 
began operations in 2009.  
It differentiated itself from contemporaneous supernova searches with its broad, untargeted survey and its scientific focus on extragalactic transients of all types \citep{Rau:09:PTFScience}.
Further aiding the success of the survey was the re-use of an existing telescope, the Palomar 48-inch Schmidt (P48), and a refurbished survey camera, the former CFHT12k \citep{Cuillandre:00:CFHT12k, Rahmer:08:PTFCamera}.
Capable, high-performance image differencing pipelines and novel machine-learning algorithms for eliminating false subtraction candidates \citep[``Real-Bogus'';][]{Bloom:12:RealBogus, Brink:13:RB2} enabled reasonable discovery purity.
Finally, the moderate depth of the PTF system (modal $R$-band limiting magnitudes of $\sim$21.0) facilitated classification spectroscopy and followup with moderate-aperture telescopes.

These factors combined to yield remarkable scientific success.  
The headlining discovery was of PTF11agg (SN2011fe), the youngest supernova Ia ever discovered \citep{Nugent:11:11klyDiscovery, Li:11:11klyProgenitor}.
Other highlights included identification of superluminous supernovae as a class \citep{Quimby:11:SuperluminousSN}, discovery of a class of Ca-rich transients in the luminosity gap between novae and supernovae \citep{Kasliwal:12:CaRich}, and discovery of a fast-fading event consistent with a proposed class of relativistic, baryon-loaded ``dirty fireballs''  \citep{Cenko:13:PTF11agg}.

For the Intermediate Palomar Transient Factory (iPTF), which began operations in 2012, two new themes emerged: exploration of cadence strategies, and rapid processing and followup of transient events.  

The collaboration conducted an internal proposal process each semester to determine the observing strategy.  
About sixty percent of the observing time was used for extragalactic transient searches, with variants including high-cadence surveys targeting local galaxies, surveys attempting to maintain strict cadence control, and surveys that obtained observations in two filters nightly.
Fifteen percent of the observation time was used for extensive variability surveys of the Galactic Plane.  
Ten percent was used for other variability-driven science, six percent for solar system surveys, and three percent for Target of Opportunity and other coordinated observations.  
The remainder was used for a narrowband ``Census of the Local Universe'' \citep{tmp_Cook:17:CLUPrelim} conducted during bright time.

Enhancements to the image processing and differencing pipelines \citep{Cao:16:iPTFPipelines} led to greater purity of the transient stream as well as much faster receipt of transient candidates (to $\sim$10 minutes, from 30--60\,minutes in PTF).  
When combined with the innovation of having live human scanning of transient candidates by partnership members in Europe, where the Palomar night overlapped with the workday, these efficiency improvements created a new capability for rapid discovery and followup.

New discoveries flowed: \citet{GalYam:14:FlashSpectroscopy} pioneered ``flash spectroscopy'', discovering that the progenitors of core collapse supernovae created fast-fading emission line signatures as they photo-ionized the circumstellar medium.
\citet{Cenko:15:iPTF14yb} provided the first discovery of a gamma-ray burst outside of the high-energy band.
\citet{Cao:15:UVPulse} triggered \textit{Swift} rapidly enough to capture the fast-fading UV pulse of an exploding white dwarf colliding with its companion star.
And \citet{Goobar:17:LensedIa} discovered the first strongly lensed Ia supernova that could be followed in real time.

Maturing pipelines and dedicated observing campaigns fed growing variable and solar system programs, yielding studies of compact binaries \citep[e.g.,][]{Bellm:16:J2129, Kupfer:17:HotSubdwarf}, asteroids \citep{Waszczak:15:AsteroidRotation,Waszczak:17:StreakingNEAs,Chang:17:SFR}, and more.

However, the increased cadence (and division of the survey into smaller sub-surveys) required a concomitant reduction in the total area of sky covered.
Accordingly, capable new sky surveys found success conducting 
wide-area surveys, usually at lower cadence: 
most notably ASAS-SN, but also including
surveys like \textit{Gaia}, ATLAS, and the PanSTARRS Survey for Transients.
The PTF surveys had maximized their output. To make further progress, a 
major enhancement in survey capability would be required.

\section{The Zwicky Transient Facility}

The ZTF project began with the simple observation that the PTF camera only filled a fraction of the available focal plane on the P48: 
PTF had an effective field of view of 7.26\,deg$^2$, while the photographic plates used by Palomar Sky Surveys covered 44\,deg$^2$, a factor of six larger.  
We sought to maximize the volumetric survey speed \citep{Bellm:16:Cadences} we could achieve on the P48.

Accordingly, we designed and built a new survey camera using sixteen 6k$\times$6k CCDs, providing an effective field of view of 47\,deg$^2$ \citep{Smith:14:ZTFSPIE, Dekany:16:ZTF}.  
However, since survey speed is also a function of image quality, throughputs (including beam obscuration), and readout and slew overheads, a range of additional optimizations helped maximize ZTF's capability. 

Modern controllers reduced readout time from the 46 seconds required for PTF to 10 seconds, providing an additional factor of two boost in survey speed.
We also upgraded the P48 telescope and dome drive motors to ensure that the telescope could slew and settle between adjacent ZTF fields within the 10 second readout time.

Maintaining PTF's moderate image quality ($\sim2^{\prime\prime}$ FWHM) 
over the full field of view required novel optics: 
in addition to the optically powered dewar window, we mounted a thin lens a few millimeters above each CCD in the faceted focal plane and added a large ``trim plate'' optic in front of the existing doublet Schmidt corrector.

Because the focal point of the P48 is within the telescope tube, minimizing beam obstruction by the camera required tight control over its cross-sectional area.
This led to our decision to install a large bi-parting shutter at the pupil of the telescope and to develop a novel filter exchanger that uses a robotic arm stowed outside of the telescope tube.  

The ZTF data processing system was based on experience gained from processing PTF \citep{Laher:14:PTFPipeline, Masci:17:PTFIDE}.  
A range of enhancements in algorithms \citep[e.g.][]{Zackay:16:ZOGY}, databases, and workflow management were required to handle the fifteen times larger data flow \citep{Laher:17:ZTFProceedings}.

\begin{figure}[b]
% \vspace*{-2.0 cm}
\begin{center}
\includegraphics[width=\textwidth]{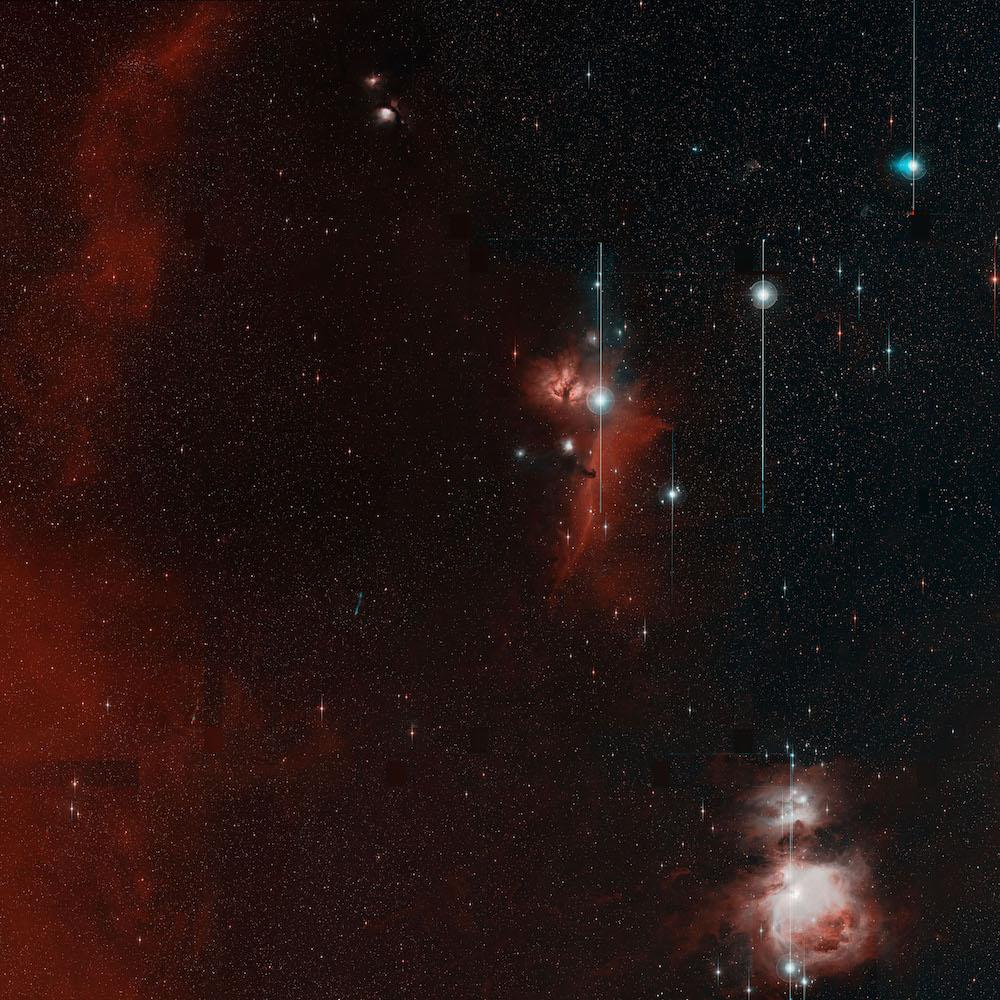}
% \vspace*{-1.0 cm}
	\caption{ZTF first light image of the constellation Orion (Caltech Optical Observatories).
	\label{fig:firstlight}}
\end{center}
\end{figure}

ZTF achieved first light in October 2017 (Figure \ref{fig:firstlight}).  
As of this writing, commissioning is ongoing; on-sky performance will be detailed in forthcoming publications (Bellm et al. in prep.).

We expect to begin formal science surveys in February 2018.
ZTF will divide its observation time between general-purpose public surveys (40\% of the observing time); surveys conducted by the ZTF partnership (40\%); and time allocated by the Caltech TAC (20\%).  For the initial 18 months of operations, ZTF will conduct two surveys in the public time: A three-night-cadence survey of the entire Northern Sky, and a nightly sweep of the visible Northern Galactic Plane ($|b| < 7^\circ$) \citep{Bellm:17:ZTF}.  
For both surveys, if a field is observed on a given night it will be observed once in $g$-band and once in $r$-band, with a gap of 30 minutes or more between observations.   
Transient alerts from the public surveys will be issued in near real-time (\S \ref{sec:alerts}). 
Annual data releases will provide access to images, catalogs, and direct imaging lightcurves.

\section{Next-generation Alert Streams with ZTF and LSST} \label{sec:alerts}

Early in the next decade, LSST \citep{2008arXiv0805.2366I} will begin delivering a massive stream of ten million nightly public alerts.
In scale, in conception, and in design, this stream will differ from the outputs of earlier time-domain surveys.
Most notably, the LSST stream will contain alerts for every source above threshold in the difference image---regardless of its physical origin or whether it has been seen before.  
There will be alert packets containing photometry for new supernovae, for instance,  but also alert packets for every detection of variable stars that pulsate stably for the entire LSST survey.
The alert stream therefore can be seen less as a channel for announcing  high-value astrophysical events and more as a means of distributing the underlying photometric database.
While individual alert packets may correspond to events of scientific importance, further classification and/or filtering will be required to identify them.  
The alert stream itself may therefore be considered a data product and thought of as a whole.  

The alert stream is naturally real-time, low-latency, and decentralized.
Because of LSST's scale, it is impractical to support followup queries to a central database in order to provide relevant contextual information needed to identify all possible events of scientific interest.   
Accordingly, LSST will provide rich alert packets that contain not just the photometry from the triggering detection and its time, but also a year's history of past observations, difference and template cutout images, and associations with objects from the yearly Data Release Processing \citep{LSE-163}.
Live (or replay) filtering of the stream replaces batch queries of a central database as the dominant means of discovery.  
The volume of alerts to be considered for any given science goal can be sharply reduced by simple filters on the basic alert packet contents, thereby facilitating more expensive crossmatch, query, or modeling operations on a reduced subset of events.

It is expected that LSST science users will identify events of interest through third-party ``community brokers'' that receive the full LSST stream and add additional value through crossmatches with external catalogs, classification, and/or filtering.  
The LSST Project will also provide a simple filtering service that operates only on the contents of the individual alert packets.
Scientific programs requiring followup observations of LSST events may benefit from integration with ``Target and Observation Managers''\footnote{also known as Marshals.} that ingest annotated and filtered streams and collate new data obtained in response to the event.

Because of the importance of third-party infrastructure in enabling the scientific success of the LSST time-domain programs, when developing ZTF we proposed to provide an ``LSST-like'' stream of alerts for the ZTF public surveys.
The ZTF stream will provide the astronomical community early experience and at smaller scale (we expect about 1 million alerts nightly from ZTF).  
To maximize fidelity with the eventual LSST stream, the ZTF alert stream will likewise contain all sources above threshold in the difference image regardless of likely astrophysical origin (subject perhaps to Real-Bogus or other quality cuts).  
We will construct rich alert packets containing recent photometric history, image cutouts, and some contextual information, including a crossmatch to the PanSTARRS-1 catalog, a star-galaxy score for the reference counterpart, and a Real-Bogus score.
We intend to forward the public streams to community brokers as well as our own simple filtering service.

We are building the ZTF Alert Distribution system using open source tools (Apache Kafka, Apache Avro, and Apache Spark) being prototyped within LSST for handling the full LSST alert stream.  
We expect the first ZTF alerts to be released in the second quarter of 2018.

\acknowledgements

ZTF and LSST are the work of many talented individuals.
I'd particularly like to acknowledge the contributions of M.~T.~Patterson, whose insights about the subtleties of streaming data I have relied heavily on in \S \ref{sec:alerts}. 

This work is based on observations obtained with the Samuel Oschin Telescope 48-inch and the 60-inch Telescope at the Palomar Observatory as part of the Zwicky Transient Facility project, a scientific collaboration among the California Institute of Technology, the Oskar Klein Centre, the Weizmann Institute of Science, the University of Maryland, the University of Washington, Deutsches Elektronen-Synchrotron, the University of Wisconsin-Milwaukee, and the TANGO Program of the University System of Taiwan. Further support is provided by the U.S. National Science Foundation under Grant No. AST-1440341.

LSST material (\S \ref{sec:alerts}) is based in part upon work supported in part by the National Science Foundation through Cooperative Agreement 1258333 managed by the Association of Universities for Research in Astronomy (AURA), and the Department of Energy under Contract No.\ DE-AC02-76SF00515 with the SLAC National Accelerator Laboratory.  Additional LSST funding comes from private donations, grants to universities, and in-kind support from LSSTC Institutional Members.

\end{document}